\PassOptionsToPackage{unicode}{hyperref}
\PassOptionsToPackage{hyphens}{url}
\PassOptionsToPackage{dvipsnames,svgnames,x11names}{xcolor}
\documentclass[
  DIV=11,
  notitlepage,
  12pt,
  a4paper,
  fleqn,
  headings=optiontohead,
  toc=indented,
  bibliography=totocnumbered,
  footinclude=false]{scrartcl}
\usepackage{xcolor}
\usepackage{amsmath,amssymb}
\usepackage{iftex}
\ifPDFTeX
  \usepackage[T1]{fontenc}
  \usepackage[utf8]{inputenc}
  \usepackage{textcomp} 
\else 
  \usepackage{unicode-math} 
  \defaultfontfeatures{Scale=MatchLowercase}
  \defaultfontfeatures[\rmfamily]{Ligatures=TeX,Scale=1}
\fi
\usepackage{lmodern}
\ifPDFTeX\else
\fi
\IfFileExists{upquote.sty}{\usepackage{upquote}}{}
\IfFileExists{microtype.sty}{
  \usepackage[]{microtype}
  \UseMicrotypeSet[protrusion]{basicmath} 
}{}
\makeatletter
\@ifundefined{KOMAClassName}{
  \IfFileExists{parskip.sty}{%
    \usepackage{parskip}
  }{
    \setlength{\parindent}{0pt}
    \setlength{\parskip}{6pt plus 2pt minus 1pt}}
}{
  \KOMAoptions{parskip=half}}
\makeatother
\makeatletter
\ifx\paragraph\undefined\else
  \let\oldparagraph\paragraph
  \renewcommand{\paragraph}{
    \@ifstar
      \xxxParagraphStar
      \xxxParagraphNoStar
  }
  \newcommand{\xxxParagraphStar}[1]{\oldparagraph*{#1}\mbox{}}
  \newcommand{\xxxParagraphNoStar}[1]{\oldparagraph{#1}\mbox{}}
\fi
\ifx\subparagraph\undefined\else
  \let\oldsubparagraph\subparagraph
  \renewcommand{\subparagraph}{
    \@ifstar
      \xxxSubParagraphStar
      \xxxSubParagraphNoStar
  }
  \newcommand{\xxxSubParagraphStar}[1]{\oldsubparagraph*{#1}\mbox{}}
  \newcommand{\xxxSubParagraphNoStar}[1]{\oldsubparagraph{#1}\mbox{}}
\fi
\makeatother

\usepackage{longtable,booktabs,array}
\usepackage{calc} 
\usepackage{etoolbox}
\makeatletter
\patchcmd\longtable{\par}{\if@noskipsec\mbox{}\fi\par}{}{}
\makeatother
\IfFileExists{footnotehyper.sty}{\usepackage{footnotehyper}}{\usepackage{footnote}}
\makesavenoteenv{longtable}
\usepackage{graphicx}
\makeatletter
\newsavebox\pandoc@box
\newcommand*\pandocbounded[1]{
  \sbox\pandoc@box{#1}%
  \Gscale@div\@tempa{\textheight}{\dimexpr\ht\pandoc@box+\dp\pandoc@box\relax}%
  \Gscale@div\@tempb{\linewidth}{\wd\pandoc@box}%
  \ifdim\@tempb\p@<\@tempa\p@\let\@tempa\@tempb\fi
  \ifdim\@tempa\p@<\p@\scalebox{\@tempa}{\usebox\pandoc@box}%
  \else\usebox{\pandoc@box}%
  \fi%
}
\def\fps@figure{htbp}
\makeatother

\NewDocumentCommand\citeproctext{}{}
\NewDocumentCommand\citeproc{mm}{%
  \begingroup\def\citeproctext{#2}\cite{#1}\endgroup}
\makeatletter
 \let\@cite@ofmt\@firstofone
 \def\@biblabel#1{}
 \def\@cite#1#2{{#1\if@tempswa , #2\fi}}
\makeatother
\newlength{\cslhangindent}
\newlength{\csllabelwidth}
\newenvironment{CSLReferences}[2] 
 {\begin{list}{}{%
  \setlength{\itemindent}{0pt}
  \setlength{\leftmargin}{0pt}
  \setlength{\parsep}{0pt}
  \ifodd #1
   \setlength{\leftmargin}{\cslhangindent}
   \setlength{\itemindent}{-1\cslhangindent}
  \fi
  \setlength{\itemsep}{#2\baselineskip}}}
 {\end{list}}
\usepackage{calc}

\providecommand{\tightlist}{%
  \setlength{\itemsep}{0pt}\setlength{\parskip}{0pt}}

\usepackage{booktabs}
\usepackage{longtable}
\usepackage{array}
\usepackage{multirow}
\usepackage{wrapfig}
\usepackage{float}
\usepackage{colortbl}
\usepackage{pdflscape}
\usepackage{tabu}
\usepackage{threeparttable}
\usepackage{threeparttablex}
\usepackage[normalem]{ulem}
\usepackage{makecell}
\usepackage{xcolor}
\usepackage[automark,headsepline]{scrlayer-scrpage}
\clearpairofpagestyles
\automark{section}
\newcommand{\shorttitle}{Dark Matter Realism}
\makeatletter
\@ifundefined{shorttitle}{\newcommand{\shorttitle}{\@title}}{}
\makeatother

\ihead{\shorttitle}
\ohead{\headmark}
\setheadsepline{0.2pt}
\setkomafont{pageheadfoot}{\small\normalfont}
\date{\normalsize\today}
\usepackage{microtype}
\PassOptionsToPackage{bold-style=ISO}{unicode-math}
\usepackage[T1]{fontenc}
\usepackage{libertinus}
\setkomafont{disposition}{\normalfont}
\setkomafont{section}{\rmfamily\large}
\setkomafont{subsection}{\rmfamily\normalsize}
\setkomafont{subsubsection}{\rmfamily\normalsize}
\setkomafont{title}{\rmfamily\LARGE}
\usepackage{mathtools}
\usepackage{mathrsfs}
\usepackage{turnstile}
\usepackage{accents}
\usepackage{upgreek}
\usepackage{logix}
\PassOptionsToPackage{dvipsnames,svgnames,x11names,table}{xcolor}
\usepackage{xcolor}
\definecolor{footrule}{HTML}{333333}
\definecolor{linkcol}{HTML}{7D3807} 
\definecolor{rulecolor}{HTML}{333333}
\definecolor{stripeOdd}{HTML}{FFFFFF}
\definecolor{stripeEven}{HTML}{EDEDED}
\renewcommand\footnoterule{\kern0pt{\color{footrule}\hrule width .4\columnwidth}\kern 6.0pt}
\usepackage{booktabs}
\usepackage{longtable}
\usepackage{tabularx}
\usepackage{array}
\usepackage{multirow}
\usepackage{transparent}
\usepackage{changepage}
\usepackage{float}
\usepackage[font=footnotesize,labelfont=bf,width=0.9\textwidth]{caption}
\usepackage{indentfirst}
\usepackage[noblocks]{authblk}

\usepackage{wrapfig}
\usepackage{tikz}
\usepackage{graphicx}
\usepackage{adjustbox}

\renewenvironment{quote}{\begin{oldquote}\small}{\end{oldquote}}
\usepackage{makecell}

\usepackage{ragged2e}
\usepackage{stmaryrd}
\newcommand{\kcrel}{\mathrel{\lower1.05ex\hbox{$\rfloor$}\mkern-5.08mu\raise1.045ex\hbox{\reflectbox{$\rceil$}}}}
\newcommand{\PHI}{ϕ }
\numberwithin{equation}{section}
\makeatletter

\newcommand*\empraise{.25ex}
\newcommand*\empkern{-.40em}

\newcommand{\emp}{\mathrel{\mathpalette\emp@inner\relax}}
\newcommand{\emp@inner}[2]{%
  \ooalign{%
    $#1\Sigma$\cr
    \hfil\kern-\empkern \raise\empraise\hbox{$#1\circ$}\hfil\cr
  }%
}
\makeatother
\makeatletter
\@ifpackageloaded{caption}{}{\usepackage{caption}}
\AtBeginDocument{%
\ifdefined\contentsname
  \renewcommand*\contentsname{Table of contents}
\else
  \newcommand\contentsname{Table of contents}
\fi
\ifdefined\listfigurename
  \renewcommand*\listfigurename{List of Figures}
\else
  \newcommand\listfigurename{List of Figures}
\fi
\ifdefined\listtablename
  \renewcommand*\listtablename{List of Tables}
\else
  \newcommand\listtablename{List of Tables}
\fi
\ifdefined\figurename
  \renewcommand*\figurename{Figure}
\else
  \newcommand\figurename{Figure}
\fi
\ifdefined\tablename
  \renewcommand*\tablename{Table}
\else
  \newcommand\tablename{Table}
\fi
}
\@ifpackageloaded{float}{}{\usepackage{float}}
\floatstyle{ruled}
\@ifundefined{c@chapter}{\newfloat{codelisting}{h}{lop}}{\newfloat{codelisting}{h}{lop}[chapter]}
\floatname{codelisting}{Listing}

\makeatother
\makeatletter
\@ifpackageloaded{caption}{}{\usepackage{caption}}
\@ifpackageloaded{subcaption}{}{\usepackage{subcaption}}
\makeatother
\usepackage{bookmark}
\IfFileExists{xurl.sty}{\usepackage{xurl}}{} 
\hypersetup{
  pdftitle={Dark Matter Realism},
  pdfauthor={Simon Allzén},
  colorlinks=true,
  linkcolor={linkcol},
  filecolor={Maroon},
  citecolor={linkcol},
  urlcolor={linkcol},
  pdfcreator={LaTeX via pandoc}}

\title{Dark Matter Realism}

\subtitle{How Referential Semantics Restricts Realism\\
in Contemporary Fundamental Physics}

  \author[123]{Simon
Allzén\thanks{simon.allzen@philosophy.su.se · \href{https://orcid.org/0000-0002-6124-8152}{ORCHID: 0000-0002-6124-8152}}}

      \affil[1]{Department of Philosophy, Stockholm University}
      \affil[2]{Institute of Physics, University of Amsterdam}
      \affil[3]{Vossius Center for the History of Humanities and
Sciences}
  
\begin{document}
\maketitle
\begin{abstract}
\noindent \small \textbf{Abstract.} Philosophers increasingly treat
semantics as decisive for realism about dark matter. In this paper, I
consider a recent proposal from Vaynberg
(\citeproc{ref-vaynberg2024realism}{2024}) anchored in the
causal-descriptive theory of reference from
(\citeproc{ref-psillos1999scientific}{Psillos, 1999},
\citeproc{ref-psillos2012causal}{2012}). I argue that the application of
Psillos' general scientific realist framework in the local context of
dark matter is misguided, partly because of the overlooked metaphysical
commitments underpinning causal-descriptivism, and partly because the
extension of `dark matter' on this account includes entities we do
\emph{not} currently consider to be dark matter, and exclude entities
that we currently consider \emph{could} be dark matter. Furthermore, I
argue that this discord between scientific realism and dark matter
should be regarded endemic in contexts where empirical evidence is
scarce: the semantic details required by the proposed scientific realism
is dependent on canonical empirical confirmation, because it is against
that background that scientific realism has been formulated and
developed.
\end{abstract}

\renewcommand*\contentsname{Contents}
{
\hypersetup{linkcolor=}
\setcounter{tocdepth}{3}
\tableofcontents
}

\begin{center}\rule{0.5\linewidth}{0.5pt}\end{center}

\section{Introduction}\label{introduction}

\noindent Philosophers engaged in exploring the marriage of scientific
realism and dark matter have, as is common in many marital contexts,
reached opposite conclusions using more or less the same semantic
apparatus. Despite their ultimate disagreement regarding realism, there
is an aspect of scientific realism that all appear to agree constitutes
a pivotal point for assessing its viability in the dark matter case: the
semantics. This paper contributes to the dark matter realism debate qua
a debate about semantics in two distinct ways.

First, I explicate the general role played by semantics in the dark
matter context and explain why it is, and should be, of particular
interest for philosophers engaged in dark matter realism. Building on
this, I examine a specific account of semantics for realism given by
Vaynberg (\citeproc{ref-vaynberg2024realism}{2024}), who employs the
causal-descriptive theory of reference
(\citeproc{ref-psillos1999scientific}{Psillos, 1999},
\citeproc{ref-psillos2012causal}{2012}) to frame a semantic
interpretation of dark matter that not only \emph{allows} for empirical
confirmation but is in fact \emph{exemplified} as such in the Bullet
Cluster observation. On this view, we are provided not only with the
semantic template for dark matter realism, but with an epistemology to
match. I do not necessarily disagree with Vaynberg's conclusion --- that
dark matter realism is justified --- but I want to raise a challenge
regarding the way he arrived there, namely by appealing to canonical
forms of scientific realism. More specifically, I argue that Psillos'
semantics, situated within his general framework of scientific realism
as it is, is unsuited as a semantic framework applied in the local
context of dark matter. This is in part because of the overlooked
metaphysical commitments under the hood of causal-descriptivism, and in
part because the conditions given for reference and object turn out to
be neither necessary nor sufficient. The extension of `dark matter' on
this account turns out to include entities we do not currently consider
to be dark matter and exclude entities that we currently consider could
be dark matter.

Second, I suggest that the failed marriage between scientific realism
and dark matter should be considered a foregone conclusion. My claim is
that the semantic elements in what appears to be the most suitable
scientific realism for dark matter is built on the following
presupposition: only canonical forms of empirical confirmation can
provide us with the causal descriptions necessary to establish that an
entity is the referent for its corresponding theoretical term. I contend
that \emph{this is a principled issue}, i.e., a built-in limit, part and
parcel of the function of scientific realism. In a sense, I am combining
the semantic lessons from Martens (\citeproc{ref-martens2022dark}{2022})
and the epistemic lessons from Allzén
(\citeproc{ref-allzen2021scientific}{2021}) to show that the
justification for dark matter realism cannot be found in classical
accounts of scientific realism. This does not, and should not, be taken
to entail or support anti-realism about dark matter, nor does it
rationally recommend suspending beliefs about the reality of dark matter
--- a staunch dark matter realist may simply conclude that if the
marriage of classic scientific realism and dark matter has failed, that
is symptomatic of the former clinging on to past scientific practice,
not the latter being unjustifiable. The main point that this paper seeks
to stress is this: it is not by accident, or because of case-specific
contingencies, that scientific realism and dark matter are at odds.
Rather, this should be an expected outcome from a doctrine the
formulation of which is so entrenched in and constructed in parallel
with the golden age of particle physics and its associated modes of
empirical confirmation. Scientific realism, understood as a child of its
time, \emph{would be} ill-equipped to accommodate theories which are so
intricately embedded in theoretical networks and so empirically
challenging as they have proven to be in 21st-century cosmology and
astrophysics. Unlike the cases that shaped traditional realism (e.g.,
well-confirmed particle physics), dark matter inhabits a regime of
persistent underdetermination and sparse direct evidence, making it a
representative case for testing the applicability of classical
scientific realism in 21st-century physics.

If, from this incompatibility, we should conclude that \emph{dark matter
realism is unjustified}, or that \emph{scientific realism is an outdated
doctrine} built on scientific practice neither desirable nor attainable
for much of current and future scientific theory, I cannot definitively
argue here. I will nevertheless suggest that the incompatibility
highlights a vulnerability in the way scientific realism (of the sort
considered) handles contemporary theory and their posited entities
which, if systematic, will render it archaic for future science.

\section{Dark matter realism and the problem of
semantics}\label{dark-matter-realism-and-the-problem-of-semantics}

\noindent In a recent contribution to the growing literature on the
philosophy of dark matter, Vaynberg
(\citeproc{ref-vaynberg2024realism}{2024}) offers an argument in support
of dark matter realism, understood in terms of an answer to the
following question: \emph{``Can one justifiably be a scientific realist
about dark matter? More specifically, does the theoretical term `dark
matter' in cosmology successfully refer to a real entity in the
world?''}\footnote{The philosophical interest in dark matter is evident
  in the growth of research in recent years: Vanderburgh
  (\citeproc{ref-vanderburgh2003dark}{2003}), Vanderburgh
  (\citeproc{ref-vanderburgh2005methodological}{2005}), Vanderburgh
  (\citeproc{ref-Vanderburgh2014-VANQPE}{2014b}), Vanderburgh
  (\citeproc{ref-vanderburgh2014interpretive}{2014a}), Sus
  (\citeproc{ref-sus2014dark}{2014}), Jacquart
  (\citeproc{ref-jacquart2021lambdacdm}{2021}), Merritt
  (\citeproc{ref-merritt2021b}{2021b}), Merritt
  (\citeproc{ref-merritt2021a}{2021a}), De Baerdemaeker \& Boyd
  (\citeproc{ref-siska2020jump}{2020}), De Baerdemaeker
  (\citeproc{ref-de2021method}{2021}), Baerdemaeker \& Dawid
  (\citeproc{ref-SiskaRichard2022mond}{2022}), Antoniou
  (\citeproc{ref-antoniou2023robustness}{2023}), Antoniou
  (\citeproc{ref-antoniou2025did}{2025}), Duerr \& Wolf
  (\citeproc{ref-WOLFDUERR20231}{2023}), Wolf \& Read
  (\citeproc{ref-wolf2025navigatingpermanentunderdeterminationdark}{2025}),
  Allzén (\citeproc{ref-allzn2024dark}{2024}), Weatherall \& Smeenk
  (\citeproc{ref-Weatherall}{forthcoming}), Martens et al.
  (\citeproc{ref-martens2022integrating}{2022}).} For anyone not already
familiar with the prerequisites of scientific realism, it may seem
strange to formulate a question about the reality of a scientific entity
as a semantic question about reference, but there are good reasons to do
so. Scientific realism is commonly glossed as the view resulting from
accepting (some version of) three core elements, listed below in order
of necessary dependence:

\begin{itemize}
\tightlist
\item
  \textbf{Metaphysical}: the world exists mind-independently.
\item
  \textbf{Semantic}: theoretic claims have truth values, and should be
  taken at face value.
\item
  \textbf{Epistemic}: science is able to identify the truth values of
  its claims.
\end{itemize}

\noindent The metaphysical element provides the ontological structures
necessary for theoretical claims to have truth values. The constitution
of the world dictates whether the theoretical claims expressed in
science, taken literally, are true or false (importantly, both
observable and unobservable entities). The epistemic element expresses
optimism that science can determine whether a theoretical claim is true
or false. If science provides sufficient reasons for the truth of a
theoretical claim, then all the entities, properties, structures, and
relations contained in that claim are (and should be) accepted as real
parts of the world. In short, scientific realism is the view one ends up
with when accepting all three dependent elements above, colloquially
expressed as the belief that our best scientific theories provide us
with knowledge about real features of the world.

Although scientific realism is classically understood as a general
perspective on science, local applications of realism to specific
theories are not exempt from dependence on these elements. Note,
however, that how the elements are realized is, in principle,
irrelevant, which implies that how one assembles the necessary
semantics, and the conceptual machinery it contains, is subject to the
theoretical aims of the realist. Once a semantic framework is
articulated, its specification --- as a consequence of elemental
dependence --- constrains the set of possible ways to specify the
epistemic conditions sufficient for empirical confirmation. Martens
(\citeproc{ref-martens2022dark}{2022}) partially draws on this
dependence in a bipartite argument against the prospects of
(contemporary) dark matter realism. According to him, the semantic
concept of dark matter has two reasonable articulations: one specific
and one generic.\footnote{My use of ``specific'' and ``generic''
  corresponds to Martens ``thick'' and ``thin''.} The specific approach
takes the concept of dark matter to be fully explicated by plugging in
the salient specifics from any currently viable dark matter candidate,
for example, a weakly interacting massive particle (WIMP). This gives us
not only a semantically meaningful concept of dark matter, but also a
discernible way to carve out the sufficient conditions for empirical
confirmation. In short, it tells us what dark matter is and how to know
when we find it. Deciding which candidate should fill the semantic gap,
however, is massively underdetermined:

\begin{quote}
the current empirical evidence strongly underdetermines which, if any,
of the vast array of completely specific mainstream candidates is to be
paired with \(\Lambda\text{CDM}\).
(\citeproc{ref-martens2022dark}{Martens, 2022, p. 4})
\end{quote}

\noindent The generic approach takes the concept of dark matter to
include only the descriptive properties shared by all currently viable
dark matter candidates, i.e., their most generic properties. Martens is
doubtful that the descriptive properties remaining after stripping the
specifics amount to a semantically meaningful concept:

\begin{quote}
The common core concept of dark `matter', especially the lower bound, is
so semantically thin, so vacuous, that it barely means anything at all.
It is simply not a rich enough concept (yet) for us to be realists about
it. (\citeproc{ref-martens2022dark}{Martens, 2022, p. 16})
\end{quote}

\noindent Here, Martens stresses the imprudence of being a realist about
the vacuous concept produced by a generic approach --- realism requires
our concept of dark matter to be specific enough to discriminate between
individual unique entities. The general reason supporting specificity is
rooted in problems of referential success. Even in cases where the
description of a scientific entity is only marginally generic --- say,
compatible with at least two distinct theoretically possible entities
--- referential success becomes trivial. The reason is that whichever of
the two entities proves to be real was somehow the one we managed to
successfully refer to all along. The damning part, of course, is that it
should not be irrelevant which of the entities actually exists for
successful reference to obtain. For example, a detective declaring that
he has successfully referred to, i.e., uniquely picked out, the
perpetrator of a crime by describing them as ``a human being'' will
surely be fired. It is a core function of our descriptions that they can
discriminate between entities sufficiently.\footnote{This is not to say
  that descriptions of entities cannot be generic, or even vacuous, at
  the conception of new theory. Such descriptions constitute the basis,
  or semantic seeds, which equips scientists with a minimal but
  necessary structure to further develop a theory. Theoretical progress
  and empirical input usually increases the specificity of the
  description, but, and this is Martens core point, realist commitment
  should not precede the point at which a theory reaches sufficient
  descriptive specificity.}

Martens' two-horned objection to semantic realism effectively terminates
the process toward dark matter realism at the semantic junction. His
clear focus on the semantic element of scientific realism explains why
Vaynberg, in response, frames the question about the justification of
dark matter realism as a question of semantics, more specifically as a
question of successful reference. In the above exposition, it is evident
that the semantic aspects of the term `dark matter' --- the
proliferation of dark matter concepts, their common descriptive core,
and the prospects of successful reference --- constitute a central issue
for dark matter realism. In the following section, I will explicate and
challenge Vaynberg's account, which I take to be the most comprehensive
and thoroughly constructed proposition in favor of dark matter realism
thus far.

\section{A semantics for dark matter
realism}\label{a-semantics-for-dark-matter-realism}

\noindent Vaynberg's semantic framing of dark matter realism, introduced
at the beginning of the previous section, persists in his more
exhaustive recipe for the view:

\begin{quote}
If we want to develop a plausible account of realism about dark matter,
then we need to stipulate a theory of reference, articulate the concept
of the theoretical entity as provided by the theory or model, and show
that the identifying properties have been empirically detected.
(\citeproc{ref-vaynberg2024realism}{Vaynberg, 2024, p. 82})
\end{quote}

\noindent The attentive reader will notice that, in addition to the
standard elements of scientific realism, a prerequisite objective is
proposed: stipulating a theory of reference. This is an advisable
measure to ensure that the semantic account one employs when dealing
with the issues raised by Martens is rooted in a vetted theory. The
theory of reference Vaynberg adopts is the causal-descriptive theory of
reference, as formulated by Psillos (\citeproc{ref-Psillos2009}{2009}).
The choice is well motivated since Psillos' version of the
causal-descriptive theory is adjusted for handling specific referential
issues of the kind found in scientific realism. According to Psillos'
theory, successful reference is achieved under the following conditions:

\vspace{0.5cm}

\begin{adjustwidth}{2em}{3em}
\begin{enumerate}
\item A term \(\mathit{t}\) refers to an entity \(\mathit{x}\) iff
      \(\mathit{x}\) satisfies the core causal description associated with \(\mathit{t}\).
\item Two terms \(t^{\ast}\) and \(\mathit{t}\) denote the same entity iff:
  \begin{enumerate}
    \item their putative referents play the same causal role with respect to a network of phenomena;
    \item the core causal description \(t^{\ast}\) takes up the kind-constitutive properties
          of the core causal description associated with \(\mathit{t}\).
  \end{enumerate}
\end{enumerate}
\end{adjustwidth}

\vspace{0.5cm}

\noindent The ``core causal description'' in (1) is constituted by the
description of the properties an entity necessarily must have in order
to cause the phenomena it was introduced to explain. In Psillos, these
descriptive properties are denoted ``kind-constitutive properties,''
since they are meant to describe the properties that constitute a
natural kind. Natural kind membership is dependent on, and determined
by, the entity possessing precisely those properties. Vaynberg adopts
Psillos' understanding of a natural kind as \emph{``a reasonable concept
meant to capture the idea that entities consisting of a set of
properties are distinct from entities that do not consist of this same
set of properties''} (\citeproc{ref-vaynberg2024realism}{Vaynberg, 2024,
p. 82}). The concept of a natural kind is thereby anchored to its
\emph{differentia}, i.e., the properties, characteristics, or any sui
generis mix thereof, which distinguish it from other kinds. Psillos
describes how all this is supposed to play out:

\begin{quote}
`Phlogiston' fails to refer because no entity has the kind-constitutive
properties attributed to phlogiston. And `oxygen' succeeds in referring,
because there is an entity with oxygen kind-constitutive properties.
{[}\ldots{]} The one and only entity to which the term refers is the
entity which is characterized by the relevant kind-constitutive
properties. (\citeproc{ref-psillos2012causal}{Psillos, 2012, p. 226})
\end{quote}

\noindent The process of identifying the relevant kind-constitutive
properties for a term is, for pragmatic reasons, necessarily
theory-dependent.\footnote{Psillos
  (\citeproc{ref-psillos2012causal}{2012, p. 226}) ``Since we have no
  theory-independent access to the kind-constitutive properties of a
  natural kind, we have to rely on theories and their causal-explanatory
  descriptions of the entities they posit.''} In Vaynberg's application
of Psillos' causal-descriptive account, the theory (or model, in this
case) most suitable to supply the kind-constitutive properties of `dark
matter' is \(\Lambda\text{CDM}\). Here follows Vaynberg's reasoning and
the subsequent stipulation of the kind-constitutive properties of dark
matter as extracted from \(\Lambda\text{CDM}\):

\begin{quote}
The mode of gravitational interaction tells us that we should expect
dark matter to interact like a collisionless fluid. This provides some
expected behavior when it comes to interactions between galactic
entities believed to possess dark matter. This also means that dark
matter must be nonbaryonic since it only interacts gravitationally and
not via electromagnetism. Together, this provides a way to articulate
\emph{the relevant kind-constitutive properties that currently form the
core causal description associated with the term `dark matter': (a)
non-baryonic and electromagnetically neutral, (b) interacts
gravitationally (acts like a collisionless fluid)}.
(\citeproc{ref-vaynberg2024realism}{Vaynberg, 2024, pp. 82--83}) {[}my
emphasis{]}
\end{quote}

\noindent Putting all this together, we arrive at the following claim,
which I take to be about the kind-constitutive properties making up the
core causal description that can satisfy the semantic element in a
scientific realism about dark matter:

\hypertarget{SDMR}{}

\begin{quote}
\emph{Semantic Dark Matter Realism (SDMR)}: The one and only entity to
which `dark matter' refers is the entity possessing the following
kind-constitutive properties: (i) non-baryonic; (ii) electromagnetically
neutral; (iii) gravitationally interactive; (iv) acts like a
collisionless fluid.\footnote{Instead of Vaynberg's (a) and (b), I
  partitioned the properties through (i) - (iv), since this more clearly
  individuates on the \emph{property level,} so is more compatible with
  the kind-constitutive framework. Note that despite adopting Psillos'
  ``one and only entity'', I take that to mean \emph{entity-kind} rather
  than entity full stop, in the sense that all members of the natural
  kind dark matter share the same set of kind-constitutive properties.
  This does not necessarily imply realism about natural kinds
  themselves.}
\end{quote}

\noindent The following evaluation of SDMR is divided into two steps.
The first scrutinizes the feasibility of applying Psillos' framework to
the dark matter case as articulated by Vaynberg. Certain theoretical
commitments embedded in Psillos' causal-descriptive theory of reference
are overlooked by Vaynberg. Explicating these embedded commitments
illuminates substantial issues when revisiting the given
kind-constitutive properties that provide the core causal description of
dark matter.

The second approach assumes, \emph{arguendo}, that issues emerging from
the first can be amended, and instead evaluates the consequences of
taking SDMR at face value. Even so, I argue, Vaynberg's semantic realism
in its current form does not deliver a semantics of dark matter that can
justify realism. The next sections expand and develop these criticisms
in the order of appearance above.

\section{Hidden metaphysics}\label{hidden-metaphysics}

\subsection{Natural kind dependency}\label{natural-kind-dependency}

\noindent Let us begin by explicating the underlying metaphysics that
underpin Psillos' causal-descriptive theory --- the metaphysics of
natural kinds. Psillos, although not explicitly defending or arguing for
any particular account of natural kinds, still makes extensive use of
concepts and ideas found in their literature. This is most evident in
his causal-descriptive theory:

\begin{quote}
Only theories can tell us in virtue of what internal properties or
mechanisms, as well as in virtue of what nomological connections, a
certain substance possesses the properties and displays the behavior it
does. Similarly, only theories can tell us in virtue of what internal
properties an item belongs to this rather than that kind. And only
theories can tell us whether a certain collection of entities, samples
or items is a candidate for a natural kind. {[}\ldots{]} The
kind-constitutive properties are those whose presence in an item makes
that item belong to a kind. I will not argue here for the existence of
natural kinds. But if there are natural kinds at all, then there are
kind-constitutive properties.
(\citeproc{ref-psillos1999scientific}{Psillos, 1999, p. 288})
\end{quote}

\noindent In later work, Psillos reaffirms his position on the
metaphysics for kind-constitutive properties, but stresses that these
properties:

\begin{quote}
are those \emph{whose presence makes} a set of objects have the same, or
sufficiently similar, \emph{manifest properties, causal behaviour}, and
causal powers. {[}\ldots{]} If we assume that natural kinds have
boundaries -- based on objective similarities and differences -- we can
see how causal descriptions succeed in fixing the reference of a term.
(\citeproc{ref-psillos2012causal}{Psillos, 2012, p. 226})
\end{quote}

\noindent Psillos' formulation of kind-constitutive properties ---
``objective similarities and differences,'' ``internal properties,''
``properties presence in an item'' --- indicates that the metaphysics at
work under the hood of his causal-descriptive theory of reference is
natural kind \emph{essentialism}. In its standard formulation,
essentialism also fits Psillos' claim that kind-constitutive properties
instantiated in an entity predetermine its causal behavior and
\emph{manifest properties}:

\begin{quote}
According to essentialism, natural kinds are groupings of entities that
share a common essence --- intrinsic properties or structure(s) uniquely
possessed by all and only members of a kind. An intrinsic property is a
property that an entity has independently of any other things, while an
extrinsic property is the one that a thing has in virtue of some
relations or interactions with other entities. The basic idea is that
the essence causes and explains all other observable shared properties
of the members of a kind and allows us to draw inductive inferences and
formulate scientific laws about them.
(\citeproc{ref-Brzovic2018-BRZNK}{Brzović, 2018})
\end{quote}

\noindent Taking stock, this means that the core causal description
meant to fix the reference of `dark matter' must necessarily consist of
kind-constitutive properties, which are the intrinsic properties
uniquely possessed by all and only members of a kind. The unique set of
intrinsic properties is the essence of dark matter, which causes and
explains the observed manifest properties shared by all instances of
dark matter, and enables inductive inferences and scientific laws.

At this point, it may be prudent to reiterate that Vaynberg accepts
Psillos' framework without modification, and that the arguments
presented in the following section are not intended as objections
against Psillos' theory in general, but rather against its application
in the dark matter case. That said, Allzén
(\citeproc{ref-allzen2021scientific}{2021}) formulates a critique
against the general feasibility of Psillos' explanationist flavor of
scientific realism using dark matter as a case study. He argues that its
explanatory power forces a Psillos-style realist to accept dark matter
realism. If my argument against Vaynberg is correct, this would
strengthen Allzén's argument against Psillos, since it would show that
his explanationist realism, and his metaphysically loaded
causal-descriptive theory of reference, are incompatible. The argument
by Allzén (\citeproc{ref-allzen2021scientific}{2021}), in tandem with
the previously explicated arguments by Martens
(\citeproc{ref-martens2022dark}{2022}), motivates Vaynberg to offer a
positive outlook on dark matter realism --- one that fully endorses a
Psillos-style semantics:

\hypertarget{V1}{}

\begin{quote}
Kind-constitutive properties are the fundamental properties the entity
must possess if it's going to play the necessary causal role and single
out the entity as being a distinct kind.
(\citeproc{ref-vaynberg2024realism}{Vaynberg, 2024, p. 82})
\end{quote}

\noindent Now that we have a better understanding of the natural kind
metaphysics that makes the causal-descriptive theory of reference tick,
let us revisit the kind-constitutive properties proposed by Vaynberg as
the core causal description of `dark matter'.

\subsection{Is dark matter a natural kind?}\label{sec-dm-natural-kind}

\noindent Vaynberg's commitment to an essentialist conception of natural
kinds is apparent in his appeal to classic essentialist schematics. We
have \emph{necessity}: in the set of kind-constitutive properties, each
property is individually necessary (all members of the kind must have
it). We have \emph{sufficiency}: taken together, the set of individually
necessary properties is sufficient to pick out the kind, since any
entity that possesses all necessary properties belongs to the kind.

\hypertarget{V2}{}

\begin{quote}
The sufficiency of the core causal description of dark matter provided
by CDM is achieved because these properties compose a set of
kind-constitutive properties that collectively make an entity belong to
a unique kind. (\citeproc{ref-vaynberg2024realism}{Vaynberg, 2024, p.
82})
\end{quote}

\noindent In the remainder of this subsection, I argue that the
properties enlisted as kind-constitutive for the core causal description
of `dark matter' are unfit for the task. They are not qualified to be
kind-constitutive and therefore cannot collectively pick out a unique
kind. The relevant question has now been reduced to: are the
kind-constitutive properties proposed by Vaynberg capable of
constituting a natural kind in the sense required by the essentialist
metaphysics on which Psillos' causal-descriptive theory of reference
depends? I argue that, viewed through the lens of essentialism, the
proposed kind-constitutive properties of SDMR appear less plausible as
candidates for a natural kind essence. Let us begin with the general
worry about properties defined by absence, exclusion, and negation that
are present in SDMR. The properties (i) non-baryonic and (ii)
electromagnetically neutral both appear as exclusionary definitions. I
use ``exclusionary'' when a property is defined by absence, exclusion,
or negation, and ``affirmative'' when defined by presence, inclusion, or
instantiation. The problem is that exclusions fail to definine intrinsic
properties:

\begin{quote}
A thing has its intrinsic properties in virtue of the way that thing
itself, and nothing else, is. {[}\ldots{]} The intrinsic properties of
something depend only on that thing. {[}\ldots{]} a sentence or
statement or proposition is entirely about something iff the intrinsic
properties of that thing suffice to settle its truth value.
(\citeproc{ref-lewis1983extrinsic}{Lewis, 1983, p. 197})
\end{quote}

\subsubsection{Non-baryonic}\label{non-baryonic}

\noindent It is questionable whether non-baryonic should be considered
an intrinsic property. It is not a property that is defined or
identified by virtue of the way dark matter is, but rather a property
defined by virtue of the way it is not. Oxford Reference defines
``non-baryonic'' as ``a hypothetical form of matter not containing
baryons --- that is, without protons or neutrons.'' The property of not
having protons and neutrons fails to qualify as an intrinsic property of
dark matter, simply because it does not say anything about what dark
matter is.

We can illuminate this with an example. Consider including ``atoms not
having 1 proton'' as a kind-constitutive property for oxygen. This
exclusionary definition only tells us that whatever oxygen is, it is not
hydrogen. It is of course \emph{true} of all members of the natural kind
oxygen that its atoms do not have 1 proton, but we cannot from that
infer what \emph{makes it true}, as per Lewis above. Having any number
of protons except 1 would make it true, which also violates uniqueness.
Our exclusionary definition effectively underdetermines which specific
number of protons makes the exclusionary definition true. Affirmative
descriptions of properties, on the other hand, face no such issues. For
instance, ``atoms having 8 protons'' \emph{makes} ``atoms not having 1
proton'' true, and from this affirmative definition we also get to
conclude that it is a candidate for uniqueness. Affirmative definitions
of properties also enable us to further specify the causal behavior
determined by an entity possessing them, since having 8 protons fixes
the electron configuration, which determines reactivity patterns,
bonding behavior, and its role in combustion.

The key point here is that only an affirmative description of a positive
attribute instantiated in dark matter can ground the truth of the
exclusionary definition `non-baryonic'. In other words, it is by virtue
of some positive, as-yet-unknown property that dark matter is
non-baryonic --- but the description of the former cannot be inferred
simply from knowing the truth of the latter.\footnote{Unless one
  provides evidence to support that baryonic/non-baryonic completely
  exhausts theoretical possibilities, making it a binary.} This
disqualifies non-baryonic as a kind-constitutive property.

\subsubsection{Electromagnetically
neutral}\label{electromagnetically-neutral}

\noindent The property of being electromagnetically neutral may be taken
as an exclusionary definition as well, since it is defined by the
absence of interaction with the electromagnetic field. We may, however,
reframe it in affirmative terms as the property of having \emph{zero
electromagnetic charge}. Yet, even when so construed, it does not seem
to qualify as a kind-constitutive property. The reason is that this
reframed definition is not universally instantiated across all dark
matter candidates, some of which are hypothesized to possess small but
non-zero charges. In addition, as a property definition it fails to
yield any positive causal entailment. Having zero electromagnetic charge
does not fix a determinate causal profile; it merely ensures that no
Coulomb attraction or repulsion, and no photon coupling, will accompany
whatever causal powers an entity \emph{does} in fact possess.
Gravitational interaction, often cited in this context, is not a
consequence of zero charge but rather a consequence of mass-energy.
Thus, zero charge operates only as an exclusionary constraint and not as
an essence-bearing property capable of individuating a natural kind.

\subsubsection{Gravitationally
interactive}\label{gravitationally-interactive}

\noindent Being gravitationally interactive is clearly formulated as an
affirmatively defined property, one that picks out an intrinsic
attribute by virtue of which an entity displays positive causal
behavior. Nevertheless, even if intrinsic, it is opaque how much this
property contributes to the essence of anything, simply because
\emph{everything} with mass/energy in the universe, as far as science
can tell, interacts gravitationally. It amounts to the trivial claim
that if dark matter is real, it has the kind-constitutive property of
being \emph{something rather than nothing}. This disqualifies (iii) as a
kind-constitutive property, because the only distinction it enforces is
that between everything and nothing.\footnote{I want to address the
  discrepancy between the phrasing ``\emph{only} interacts
  gravitationally'' in the quote from Vaynberg and the subsequent
  condition ``interacts gravitationally'' in the core causal
  description. This difference materially impacts the range of possible
  entities that can be logically included. I take this discrepancy to be
  a clerical oversight, based on assuming that the sentence containing
  ``only interacts gravitationally'' most likely intended gravitational
  interaction to serve as a contrast class to electromagnetic
  interaction. Nevertheless, I shall cover this version as well, just in
  case. When ``only'' is included, we introduce a significant constraint
  on possible dark matter candidates, by definition excluding a range of
  dark matter candidates not yet empirically ruled out: WIMPs couple via
  the weak force and the Z/Higgs, Scalar singlet interacts via Higgs
  exchange, Fermion singlet to the Higgs and the weak force by mixing,
  minimal dark-matter multiplets (\(\mathrm{SU} (2)_L\)) couple to the
  weak force as does the lightest Kaluza-Klein particle. Although the
  inclusion of a restrictive modifier acts as a clear differentia for
  natural kindhood, it obviously excludes too much.}

\subsubsection{Collisionless fluid}\label{collisionless-fluid}

\noindent Acting like a collisionless fluid perhaps most obviously fails
to qualify as an intrinsic property, since collisions by definition are
extrinsic behavior. Moreover, we once again see the exclusionary
addition of \emph{not} colliding with other entities (or itself). This
yields a relational, extrinsic property, characterized by its inability
to be part of relational causal behavior. Nevertheless, this is perhaps
the most interesting property in the set since it is an emergent
property sensitive to both environmental conditions and scale. At
sufficiently large scales, galaxies exemplify this property, despite the
fact that part of their total mass is baryonic, which, on sufficiently
small scales, does not itself exemplify this property. This makes it
difficult to assess it as a property an entity possesses, rather than as
an emergent property manifesting in systems under certain conditions. At
any rate, (iv) should be disqualified as a kind-constitutive property.

In summary, all of the properties enlisted to serve as the
kind-constitutive basis for the core causal description of `dark matter'
turn out to be incompatible with the very metaphysical framework that
renders the notion of kind-constitutive properties intelligible in the
first place. Essentialism about natural kinds may well retain its
precision and legitimacy within the general contours of Psillos'
scientific realism (although see the remark regarding the paper by
Allzén (\citeproc{ref-allzen2021scientific}{2021})), but it fails when
confronted with cases at the margins, such as dark matter. The
exclusionary, relational, and ubiquitous character of the proposed
properties leaves us without a positive, intrinsic attribute capable of
sustaining kind-constitutiveness. If this is still not convincing, we
can consider the consequences of accepting Vaynberg's proposed
properties as the core causal description of `dark matter' at face
value.

\section{A model-space for dark
phenomena}\label{a-model-space-for-dark-phenomena}

\noindent For the sake of argument, let us ignore the issues outlined in
the previous section and assume that the properties proposed as
kind-constitutive for a core causal description are qualified as such by
some metaphysics or other. Taking \hyperlink{SDMR}{SDMR} at face value
implies that for any entity, if that entity possesses the
kind-constitutive properties (i)--(iv), it is a member of the natural
kind dark matter. This follows from the claim that the relevant
properties are jointly sufficient:

\hypertarget{V3}{}

\begin{quote}
Since there is not an already existing, empirically confirmed entity
that satisfies the kind-constitutive properties associated with the core
causal description, whatever satisfies the reference of `dark matter'
will belong to this kind. This set of identifying properties will be
consistent with future discoveries of dark matter particles no matter
how many different types of these particles are found.
(\citeproc{ref-vaynberg2024realism}{Vaynberg, 2024, p. 82})
\end{quote}

\noindent The set of statements jointly offered in support of, and as a
definition of, dark matter can be summarized as follows. The term dark
matter refers to the members of a natural kind. Membership in the
natural kind is restricted to entities instantiating a set of properties
(i)--(iv). Properties (i)--(iv) satisfy the core causal description from
\(\Lambda\text{CDM}\). All present or future entities that satisfy the
core causal description by instantiating properties (i)--(iv) will be
members of the natural kind, and thus what the term dark matter refers
to. At present, there is no empirically confirmed entity that satisfies
the core causal description by instantiating properties (i)--(iv) and
thereby serves as the referent of \emph{dark matter}.

In order to gauge the implications of SDMR, it is useful to think about
the space of possible models of dark matter. Below is a compact,
many-sorted first-order language framework that we can call
\(\mathcal{L}\), that will help us articulate a model-space for dark
matter models.

\begin{table}[H]
\centering
\caption{$\mathcal{L}$: signature}
\label{tab:signature-extension}
\begingroup\small
\rowcolors{2}{stripeOdd}{stripeEven}
\begin{tabularx}{0.85\textwidth}{
  >{\raggedleft\arraybackslash}p{0.1\textwidth}
  >{\raggedright\arraybackslash}p{0.31\textwidth}
  X
}
\toprule
\rowcolor{white}
\textbf{Symbol} & \textbf{Term} & \textbf{Object} \\
\midrule
\PHI & variable & network of phenomena \\
$\varphi$ & variable & phenomena \\
$x$ & variable & entities \\
$s, o$ & constants & modes of $x$ (structure or object) \\
$\alpha$ & variable & ontological individual \\
$P$ & unary predicate & properties \\
$dm$ & individual constant & theoretical term ``dark matter'' \\
$\mathcal{m}$ & variable & model \\
$\mathbb{M}$ & set & over models \\
$\mathrm{E},\mathrm{C},\mathrm{D}$ & ternary predicate & takes $\mathcal{m}, x, \varphi$ \\
$\emp$ & unary predicate & empirically confirmed \\
$\mathrm{R}(t,\alpha)$ & binary predicate & term $t$ refers to $\alpha$ \\
$\WhiteDiamond P(x)$ & modal operator & possibility \\
\bottomrule
\end{tabularx}
\endgroup
\end{table}

\noindent First, we can define the \emph{explanandum} for which the term
`dark matter' is taken to be the \emph{explanans}, as well as the cause.
We remain theory-neutral and simply define it in terms of unexplained
phenomena for which dark matter has been invoked as an explanation, or
in terms of `dark phenomena':

\begin{quote}
The fact of the matter is that the conjunction of the assumptions i)
\href{https://www.sciencedirect.com/topics/physics-and-astronomy/general-relativity}{General
Relativity} (GR), and Newtonian Gravity as its non-relativistic limit,
is the correct theory of gravity (and spacetime) and ii) most of the
matter in the universe is luminous baryonic matter (the stuff that stars
and planets consist of), leads to predictions that have been falsified
by observations
(\href{https://www.sciencedirect.com/science/article/pii/S135521982030109X\#bib2}{Bertone
\& Hooper, 2018};
\href{https://www.sciencedirect.com/science/article/pii/S135521982030109X\#bib1}{Sanders,
2010}). We will call these observations `dark phenomena' or `dark
discrepancies'. (\citeproc{ref-MARTENS2020237}{Martens \& Lehmkuhl,
2020, p. 2})
\end{quote}

\noindent Let \(\text{\PHI}\!_{\mathrm{dark}}\) contain exactly those
phenomena the cause and explanation of which are attributed to dark
matter, the ``dark phenomena'' of Martens \& Lehmkuhl
(\citeproc{ref-MARTENS2020237}{2020}).

\begin{align}
\text{\textsc{dark matter phenomena:}}& & &\text{\PHI}\!_{\mathrm{dark}} \mathrel{\mathop:}= \bigl\{\varphi : \varphi_{dark} \bigr\}
\end{align}

\noindent We have two individual constants \(o\) and \(s\) to represent
the two modes of ontology represented in dark matter models such that:

\[
o \neq s \quad \text{and} \quad \forall x\,(x=o \ \lor\ x=s)\quad 
\text{(here $x$ ranges over \emph{modes} when used in $\mathrm{E},\mathrm{C},\mathrm{D}$)}
\]

\noindent This ensures that even though the variable \(x\) is present,
it is taken to be either an ontological object (\(o\)) or an ontological
structure (\(s\)) when used in the context of the predicates
\(\mathrm{C}, \mathrm{E}, \mathrm{D}\). These predicates should be
interpreted as expressing that the model ascribes a cause for the
phenomena, an explanation of the phenomena in terms of the cause, and a
description of the entity causing the phenomena:

\begin{itemize}
\tightlist
\item
  \(\mathrm{E}(\mathcal m, x, \varphi)\): in \(\mathcal m\), \(x\) is a
  cause of \(\varphi\),
\item
  \(\mathrm{C}(\mathcal m, x, \varphi)\): \(\mathcal m\) explains
  \(\varphi\) with \(x\), and;
\item
  \(\mathrm{D}(\mathcal m, x, \varphi)\): \(\mathcal m\) describes \(x\)
  \emph{qua} cause of \(\varphi\)
\end{itemize}

\noindent We can then define the model-space for dark matter models as:

\begin{align}
\quad\mathbb{M}\;\Defn\;\Bigl\{\;\mathcal m \colon\forall\varphi\in\text{\PHI}\!_{\mathrm{dark}}\exists x \  \Bigl(\mathrm{E}(\mathcal m,x,\varphi)\ \wedge\ \mathrm{C}(\mathcal m,x,\varphi)\ \wedge\ \mathrm{D}(\mathcal m,x,\varphi)\Bigr)\Bigr\}
\end{align}

\subsection{Regions and boundaries}\label{regions-and-boundaries}

\noindent We can partition model-space to distinguish a subset of
\(\mathbb{M}\) in which all models are restricted with respect to
ontological \emph{objects}, i.e., models that add new ontological kinds
to explain the phenomena. These would be models typically containing a
new particle, field, or other novel ontic object, taken to cause and
explain \(\varphi\), which the model provides a description of. We
define this subset, \(\mathbb O\), as:

\begin{align}
\quad\mathbb O\;\Defn\;\Bigl\{\;\mathcal m\colon\forall\varphi\in\text{\PHI}\!_{\mathrm{dark}} \exists x \ \Bigl(x = o \land \mathrm{E}(\mathcal m, x, \varphi)\ \wedge\ \mathrm{C}(\mathcal m,x,\varphi)\ \wedge\ \mathrm{D}(\mathcal m,x,\varphi)\ \Bigr)\Bigr\}
\end{align}

\noindent Mirroring the same reasoning gets us a partition of
\emph{structure}-restricted models, \(\mathbb{S}\), i.e., models that do
\emph{not} add new ontological kinds to explain the phenomena, but
rather make the case that the phenomena can be explained by a
reconfiguration of already accepted entities and their properties
(\(s\)). These would be models typically attempting to reconceptualize
the description of gravity such that the resulting gravitational
dynamics is taken to cause and explain \(\varphi\).

\begin{align}
\quad\mathbb S\;\Defn\;\Bigl\{\;\mathcal m\colon\forall\varphi\in\text{\PHI}\!_{\mathrm{dark}} \exists x \,\Big( x = s \land \mathrm{E}(\mathcal m, x, \varphi)\ \wedge\ \mathrm{C}(\mathcal m,x,\varphi)\ \wedge\ \mathrm{D}(\mathcal m,x,\varphi)\Bigr)\Bigr\}
\end{align}

\noindent We define the current catalog of known, empirically viable
models as a subset of object- and structure-based models,
\(\mathbb K \subset (\mathbb O \cup \mathbb S)\). For completion, we can
define a space in \(\mathbb{M}\) designated to unknown, or unconceived,
alternative models, i.e., models that give possible explanations and
causes for \(\text{\PHI}\!_{\mathrm{dark}}\) described by some
yet-to-be-considered physics, as:

\begin{align}
\quad &\mathbb U\;\Defn\;\{\,\mathcal m\in\mathbb M \colon \mathcal m\notin\mathbb K\,\}
\end{align}

\noindent This makes the model-space completely constituted by their
partitions in the following way:

\begin{align}
\quad\mathbb M\;=  \mathbb O \cup \mathbb S \cup \mathbb U 
\end{align}

\noindent According to Vaynberg, any future detection of \emph{actual}
dark matter will show that, whatever model is true of actual dark
matter, it will contain the kind-constitutive properties in
\(\mathbb{VDM}\). That implies a sharp boundary for what a possible
model of dark phenomena in model-space can be. Let's see how the space
of possible dark matter models looks under Vaynberg's constraints. We
can abbreviate the four kind-constitutive properties in SMDR as follows:

\[
\quad  \mathbb{VDM} \colon 
\left\{\;
\begin{aligned}
\operatorname{NB} =&  \text{ Non-baryonic} \\
\operatorname{EN} =&  \text{ Electromagnetically neutral} \\
\operatorname{GR} =&  \text{ Gravitationally interacting} \\
\operatorname{CF} =&  \text{ Acts like a collisionless fluid}
\end{aligned}
\;\right\}
\]

\noindent We take the joint instantiation of these properties in some
ontological object to be \(\mathbb{VDM} (\alpha)\). This encodes, or
imparts, the joint property constraints
(\(\operatorname{NB}, \operatorname{EN}, \operatorname{GR}, \operatorname{CF}\))
on the \emph{object} \(\alpha\) to which the term ``dark matter'' refers
(object because the kind-constitutive properties exclude models of
modified gravity --- which hold that the referent of `dark matter' is
void). Vaynberg's \(\mathbb{VDM} (\alpha)\) gives us a boundary
condition for what dark matter \emph{could possibly be}:

\begin{align}
\text{\textsc{vaynberg constraint:}}& & &\mathbb V\ \Defn\ \{\ \alpha \colon \mathbb{VDM} (\alpha) \}
\end{align}

\begin{figure}[H]

\centering{

\includegraphics[width=1\linewidth,height=\textheight,keepaspectratio]{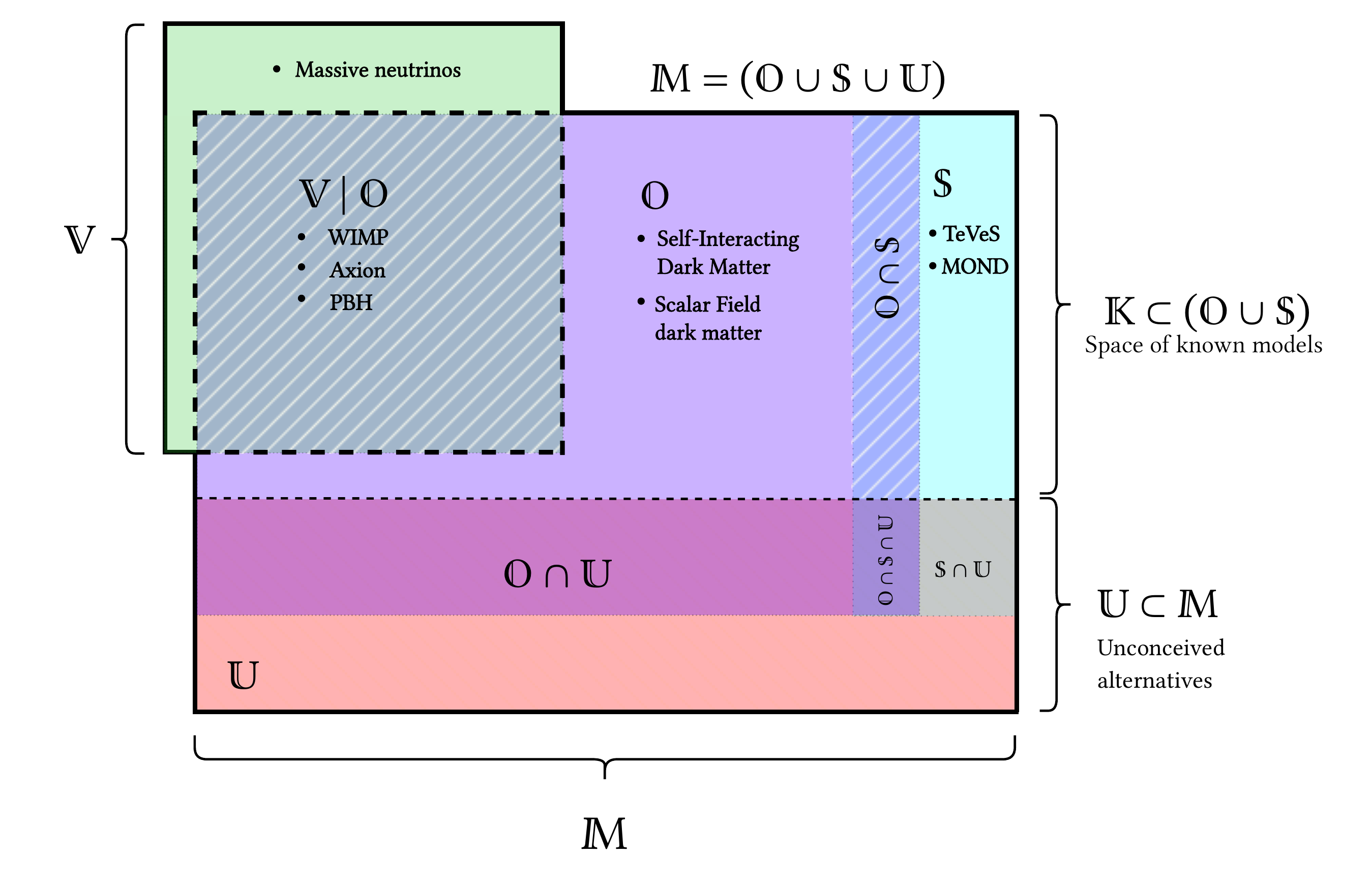}

}

\caption{\label{fig-modelspace}Model-space. Y axis is split between
known and unknown models, the X axis is split between object and
structure based models. The shaded area is the Vaynberg constraint
\(\mathbb V\).}

\end{figure}

\noindent This is the space of entities that could possibly be the
referent of `dark matter' --- a constraint for what is ontologically
admissible for a model --- as predicted by the claim that any future
detection of dark matter, whatever it may amount to, will at least
satisfy the Vaynberg constraint. We can think of this as a prerequisite
condition for membership in the dark matter kind. Models in
\(\mathbb O\) compatible with some \(\alpha \in \mathbb V\) are inside
the Vaynberg boundary; pure \(\mathbb S\) models fail the
object-requirement. \(\mathbb V\) constrains which entities can be
realized by models in \(\mathbb M\); some entities admissible by
\(\mathbb V\) may have no current model in \(\mathbb K\). To make the
model-space partitions, as well as the Vaynberg constraint, more
intuitive, we can use Figure~\ref{fig-modelspace} as a heuristic device.
The idea is to get a spatial map of ``where models live'' based on the
partitions. \(\mathbb V\) is not a strict part of model-space, but is
superimposed on top of it, highlighting the region where its constraints
correspond to a model specification.

\section{The referential failure of `dark
matter'}\label{the-referential-failure-of-dark-matter}

\noindent The plotted models in the diagram below foreshadow the central
points I am about to make in this section, in the sense that we see the
\(\mathbb V\) boundary of admissible objects extend beyond model space,
and additionally cut right through the space of known models. This
section clarifies why, and provides a diagnostic for SDMR. I identify
two types of \emph{semantic mistakes}, showing that the
kind-constitutive properties are neither necessary nor sufficient for
the intended purpose.

\subsection{Semantic mistake: type a}\label{semantic-mistake-type-a}

\noindent One type of semantic mistake shows that the kind-constitutive
properties are not \emph{necessary} for securing the referent. We can
say that it is possible that an individual object \(\alpha\) is in a
causal-explanatory model for phenomena
\(\varphi \in \text{\PHI}\!_{\mathrm{dark}}\), and `dark matter' refers
to \(\alpha\) (\(\mathrm R(dm, \alpha)\)). But if \(\alpha\) does not
instantiate the joint property constraints in \(\mathbb{VDM}\), then the
kind-constitutive properties are not necessary for securing the referent
of `dark matter'. This is a failure of \emph{necessity}.

\begin{align}
\text{\textsc{type a ($\neg$ necessary):}}& & &\WhiteDiamond \exists \alpha \; \Big(
  \mathrm{R}(dm,\alpha) \,
  \wedge
  \neg \mathbb{VDM}(\alpha) \Big)
\end{align}

\noindent Vaynberg tells us that the ``mode of gravitational interaction
tells us to expect dark matter to act like a collisionless fluid.'' This
expectation comes from galaxy-scale observations and cosmological
modeling under \(\Lambda\text{CDM}\). This implies no significant
non-gravitational self-scattering that alters those dynamics in the
relevant regimes. So, in this context, ``collisionless'' isn't a mere
approximation --- it's a kind-constitutive property that, if violated,
changes the kind-membership. The \(\Lambda\text{CDM}\) role is preserved
across several live options that violate \(\operatorname{CF}\)
condition. Therefore, \(\operatorname{CF}\) \emph{cannot} be necessary
for kind-membership on a causal-descriptive reading.

\begin{enumerate}
\def\labelenumi{(\arabic{enumi})}
\item
  \textbf{Self-interacting dark matter}\\
  \noindent Cold halos with elastic self-scattering produce cores and
  shape/abundance effects; a velocity-dependent cross section aligns
  dwarfs/galaxies vs clusters. Example:
  \(\sigma/m \sim 0.1\text{–}1\ \mathrm{cm}^2\mathrm{g}^{-1}\) at
  \(v\lesssim300 \mathrm{km\ s^{-1}}\), dropping to \(\lesssim 0.1\) at
  cluster scales. This is explicitly non-collisionless while preserving
  the \(\Lambda\text{CDM}\) gravitational function. See Spergel \&
  Steinhardt (\citeproc{ref-SpergelSteinhardt2000}{2000}), Tulin \& Yu
  (\citeproc{ref-TulinYu2018}{2018})
\item
  \textbf{Dissipative / double-disk dark matter}\\
  \noindent A dark \(U(1)\) with a (possibly light) dark photon enables
  radiative cooling in a subcomponent of dark matter, forming dark
  disks/compact substructure while the bulk remains cold on large
  scales. Example: kinetic mixing \(\varepsilon\lesssim10^{-9}\)
  (model-dependent); a sub-dominant fraction \(f_D\lesssim\text{few}%
  \) is enough to alter inner-galaxy dynamics. Again, not collisionless.
  See Fan et al. (\citeproc{ref-FanEtAl2013}{2013})
\item
  \textbf{Mirror dark matter}\\
  \noindent An exact standard model copy (\(e',p',H',\dots\)) with
  gravity and tiny photon--mirror-photon mixing. Mirror electromagnetism
  makes the sector dissipative, permitting cooling, disks, and
  distinctive halo thermodynamics --- not collisionless --- yet the same
  gravitational role is played. Example: \(\varepsilon\sim10^{-9}\)
  appears in halo and supernova-heating phenomenology. See Foot
  (\citeproc{ref-Foot2014}{2014})
\item
  \textbf{Superfluid dark matter}\\
  \noindent Strongly interacting, light bosons condense in galactic
  interiors; phonon excitations mediate an additional force reproducing
  MOND-like scaling while retaining \(\Lambda\text{CDM}\) on
  cosmological scales. The inner halo is a fluid with non-negligible
  self-interaction --- not collisionless --- yet still fits the
  gravitational brief. Example: superfluid phase in galaxies; normal
  phase in clusters (velocity/temperature dependent). Berezhiani \&
  Khoury (\citeproc{ref-BerezhianiKhoury2015}{2015}), Khoury
  (\citeproc{ref-Khoury2016}{2016})
\item
  \textbf{Strongly interacting massive particles}\\
  \noindent Relic abundance arises via \(3 \to 2\) processes in a
  strongly-coupled dark sector; the same dynamics generically yield
  large elastic self-interactions in the core-forming window. Example:
  \(m\sim10\text{–}100\ \mathrm{MeV}\); \(\sigma/m\) often in the
  \(0.1\text{–}10\ \mathrm{cm}^2\mathrm{g}^{-1}\) regime
  (model-dependent). See Hochberg et al.
  (\citeproc{ref-HochbergEtAl2014}{2014})
\end{enumerate}

\noindent In all of the above models, the relevant explanatory
gravitational role is preserved while \emph{collisionless} is
systematically violated (elastic self-scattering; radiative cooling;
phase structure; strong number-changing dynamics). Therefore,
\(\operatorname{CF}\) is \emph{not} necessary for the causal-explanatory
function \(\Lambda\text{CDM}\) assigns to dark matter. Any semantics
that includes it in the kind-constitutive core will misclassify live
realizers.

\subsection{Semantic mistake: type b}\label{semantic-mistake-type-b}

\noindent Another type of semantic mistake shows that the
kind-constitutive properties are not \emph{sufficient} for securing the
referent. We can say that when an individual object \(\alpha\)
instantiates the joint property constraints in \(\mathbb{VDM}\), then
`dark matter' refers to \(\alpha\) (\(\mathrm R(dm, \alpha)\)). But if
\(\alpha\) is a known empirically confirmed entity \emph{not} taken to
be dark matter, then \(\mathbb{VDM}\) is not sufficient to secure the
referent of `dark matter'. This is a failure of \emph{sufficiency}.

\begin{align}
\text{\textsc{type b ($\neg$ sufficient):}}& & &\exists \alpha \, \Big(\emp\!(\alpha) \wedge
\mathbb{VDM}(\alpha) \wedge \neg \mathrm{R}(dm,\alpha) \Big)
\end{align}

\noindent The second failure shows that the kind-constitutive
constraints are not sufficient to secure reference. There are ontic
candidates that do instantiate the joint property constraints in
\(\mathbb{VDM}\) but to which `dark matter' does not (even possibly)
refer, because they fail to cause/explain the relevant phenomena in
\(\text{\PHI}\!_{\mathrm{dark}}\).

\begin{enumerate}
\def\labelenumi{(\arabic{enumi})}
\tightlist
\item
  \textbf{Standard model neutrinos}\\
  \noindent Let \(\alpha=\nu\) denote the cosmic background of massive,
  active neutrinos. Neutrinos are non-baryonic \((\operatorname{NB})\),
  electromagnetically neutral \((\operatorname{EN})\), gravitationally
  interacting \((\operatorname{GR})\), and (for structure formation)
  effectively a collisionless fluid \((\operatorname{CF})\). Hence,
  \(\nu\) satisfies the \emph{joint constraints} in
  \(\mathbb{VDM}(\nu)\).\footnote{On the cosmology of massive neutrinos,
    see Lesgourgues \& Pastor
    (\citeproc{ref-LesgourguesPastor2006}{2006}), for the parameter
    imprint and suppression of small-scale power, see Hu et al.
    (\citeproc{ref-HuEisensteinTegmark1998}{1998}).}
\end{enumerate}

\noindent Yet \(\nu\) fails the causal--explanatory role that
\(\Lambda\text{CDM}\) assigns to dark matter across
\(\varphi \in \text{\PHI}\!_{\mathrm{dark}}\) (e.g., seeding/maintaining
small-scale structure, halo formation). The reason is
\emph{free-streaming}: relativistic (or semi-relativistic) neutrinos
erase perturbations below a characteristic scale, suppressing the matter
power spectrum at \(k \gg k_\mathrm{fs}\) roughly in proportion to the
neutrino fraction \(f_\nu=\Omega_\nu/\Omega_m\):

\begin{align}
\frac{\Delta P}{P}\,\simeq\,-\,8\,f_\nu& &\text{and}& &\Omega_{\,\nu}\, h^2 \;=\; \frac{\Sigma m_{\,\nu}}{93.14\ \mathrm{eV}}
\end{align}

\noindent Empirically, the observed abundance of small-scale structure
and CMB/large scale structure constraints on \(\Sigma m\_\nu\) jointly
rule out neutrinos as the \emph{dominant} dark matter component; they
cannot realize the required gravitational-explanatory role in
\(\text{\PHI}\!_{\mathrm{dark}}\).\footnote{For contemporary
  cosmological bounds and the associated structure-growth implications,
  see Planck Collaboration (\citeproc{ref-Planck2018Params}{2020}).}

There exists an \(\alpha(\nu)\) such that \(\mathbb{VDM}(\alpha)\) holds
while \(\neg\mathrm{R}(dm,\alpha)\) also holds. Therefore,
\(\mathbb{VDM}\) is not sufficient for referential success on a
causal-descriptive semantics. Strengthening \(\mathbb{VDM}\) by adding
ad hoc clauses (e.g., ``coldness,'' or a bound on free-streaming) simply
begs the question, since those clauses are \emph{fixed by} the very
explanatory successes at issue. In conclusion, the kind-constitutive
properties in \(\mathbb{VDM}\) fail to be either necessary or sufficient
for securing the referent of `dark matter' on a causal-descriptive
semantics. This renders SDMR semantically defective in the dark matter
case, and thereby breaks the case for realism about dark matter.

\section{Scientific realism and empirical
scarcity}\label{scientific-realism-and-empirical-scarcity}

\noindent Above, I argued that applying scientific realism to the dark
matter case as per Vaynberg (\citeproc{ref-vaynberg2024realism}{2024})
is plagued by incompatibility in both metaphysics and semantics. It is
my suspicion that this incompatibility is not accidental, nor merely
attributable to idiosyncrasies of the dark matter context. Rather, we
should view it as endemic to scientific realism in an era in which
theory in cosmology, astrophysics, astronomy, and particle physics
proves difficult to couple with empirical evidence \emph{of the sort
that scientific realism presupposes}. This closing section expands the
rationale for this worry by showing how it manifests in the dark matter
case and why there is reason to suspect that it extends, more generally,
to theoretical environments characterized by empirical scarcity.

\subsection{Dark matter, specificity, and
truth-preservation}\label{dark-matter-specificity-and-truth-preservation}

\noindent As we have seen, employing the (causal-descriptive) semantics
of scientific realism in a case like dark matter is not straightforward.
I have explicated two reasons:

\begin{enumerate}
\def\labelenumi{\Alph{enumi}.}
\item
  \textbf{\textbar{}} \emph{Metaphysics}\\
  Causal-descriptive semantics rely on a metaphysics of natural kinds as
  essences, explicated as the \emph{intrinsic} properties of an object
  which, taken together, constitute a unique kind.
\item
  \textbf{\textbar{}} \emph{Semantic errors}\\
  Accepting the causal-descriptive semantics for dark matter at face
  value showed that its extension failed to contain a meaningful set of
  entities and structures featured in models currently considered live
  options.
\end{enumerate}

\noindent There are reasons to think that B fails for the same
underlying reason as A: the semantics of scientific realism require a
level of empirical detail that is not available in the dark matter case.
The discussion of intrinsic properties in
Section~\ref{sec-dm-natural-kind} showed that the metaphysics of natural
kinds is ill-suited here in its current state because we lack the
empirical resolution needed to identify a unique set of intrinsic
properties. We do not know what makes it true that dark matter is, for
instance, non-baryonic. Non-baryonic is best described as a determinable
property or a genus-level class of properties, like \emph{color} or
\emph{shape}.\footnote{A determinable property is a general property
  that can be realized by multiple specific properties. For example, the
  property of being a ``color'' is a determinable property because it
  can be realized by specific colors like red, blue, or green. In
  contrast, a determinate property is a specific property that uniquely
  identifies an entity. For example, the property of being ``red'' is a
  determinate property because it uniquely identifies a specific color.
  See Wilson (\citeproc{ref-sep-determinate-determinables}{2023}).} Such
properties can be made more precise by specifying a \emph{determinate}
property or a \emph{species} of non-baryonic matter --- analogous to
specifying a determinate red for the determinable
\emph{color}.\footnote{Note that \emph{red} is also the determinable of
  properties of increasing resolution: crimson, scarlet, vermilion,
  burgundy, etc.} The semantics of scientific realism, as currently
formulated, operate on the latter, not the former.

The problem can be expressed in terms of truth-conditions using the
concept \emph{hyperintensionality}, used to classify sensitivity to
differences in meaning that don't change truth-conditions.\footnote{``A
  \emph{hyperintensional} concept draws a distinction between
  necessarily equivalent contents. If the concept is expressed by an
  operator, \(H\), then \(H\) is hyperintensional insofar as \(HA\) and
  \(HB\) can differ in truth value in spite of \(A\) and \(B\)'s being
  necessarily equivalent'' Berto \& Nolan
  (\citeproc{ref-sep-hyperintensionality}{2025})} Even when different
models agree on determinables (e.g., NB, EN, GR, CF), they may disagree
in their truth-makers --- the fine-grained grounds that make those
determinables true. WIMPs and axions are both non-baryonic
extensionally, yet what makes this true differs hyperintensionally ---
WIMPs being neutralinos and QCD axions being Peccei--Quinn (pseudo)
Nambu-Goldstone Bosons (PBHs are non-baryonic for a third, non-particle,
reason). Consider:

\begin{enumerate}
\def\labelenumi{\alph{enumi}.}
\tightlist
\item
  Dark matter is non-baryonic
\item
  Dark matter is a QCD Axion
\end{enumerate}

\noindent According to Vaynberg, (a) is necessarily true. If we suppose
that (b) is true, and ``QCD axion'' is a proper name, then (b) is
necessarily true. This renders (a) and (b) necessarily equivalent, such
that the truth of (a) is preserved when we substitute the predicate
``non-baryonic'' for ``QCD axion.'' However, we can know (a) without
knowing (b). The reason (a) appears robust, and why it features in
\(\mathbb{VDM}\), is that its truth value is preserved under model
substitution --- i.e., it is (supposed to be) model-invariant. This can
be thought of as the truth-equivalence between (a) and a disjunction of
predicates:

\begin{enumerate}
\def\labelenumi{\alph{enumi}.}
\setcounter{enumi}{2}
\tightlist
\item
  Dark matter is a WIMP or dark matter is a QCD axion or dark matter is
  a PBH \(\dots\)
\end{enumerate}

\noindent The truth of (a) is preserved when substituting its predicate
with the disjunction of predicates in (c). However, the truth of the
individual disjuncts --- the predicate-propositions in (c) --- is not
preserved if we substitute the predicate in one disjunct with a
predicate from another. In the analogue, ``x has a color'' is true if
``x is red or x is blue or x is green \(\dots\)'' is true, but ``x is
red'' does not remain true if we substitute ``red'' for ``blue'' or
``green.'' The truth of the determinable is preserved under substitution
of the disjunction of determinates, but the truth of the determinates is
not preserved under substitution of one another. This is a
hyperintensional distinction, and it makes all the difference for
scientific realism.

One reason is that truth-conditions vary with the level of precision in
the description of the referent. The more precise the description, the
more constrained the set of possible realizers; the less precise the
description, the less constrained the set of realizers. This is a
problem for scientific realism because it relies on a
\emph{causal-descriptive} semantics that requires a unique set of
intrinsic properties to fix reference. If we only have a determinable,
then we cannot uniquely fix reference because there are many possible
realizers. For example, ``dark matter is non-baryonic'' can be made true
by many different types of non-baryonic matter, such as WIMPs, axions,
or primordial black holes. The truth of the statement is invariant with
respect to model choice at this level of discrimination. But these
models vary in what makes it true that dark matter is non-baryonic. As
in the proton-number case from Section~\ref{sec-dm-natural-kind}, unless
we provide the determinate property, we cannot provide the intrinsic
properties necessary for scientific realism to apply. It is not enough
to describe an entity by its determinables. We need to know what
\emph{makes it true} that the entity has those properties. It may be a
true description of an entity that it has both determinables
\emph{color} and \emph{shape}, but realism is apt only when we can say
that an entity is red and round, rather than, say, blue and square.

\begin{figure}[h!]
\centering
\begin{minipage}{0.43\textwidth}
\raggedleft
\footnotesize
\renewcommand{\arraystretch}{1.415}
\rowcolors{2}{stripeEven}{stripeOdd}
\begin{tabularx}{\linewidth}{@{}>{\raggedleft\arraybackslash}X@{}}
\toprule
WIMPs (neutralino) \\
\midrule
SUSY electroweak state; no color or quark \\
$q_{EM}=0$; only weak or Higgs portals \\
Massive particle; $T_{\mu\nu}\neq 0$ \\
Tiny self-scatter $\sigma_{\text{self}}/m\!\ll\!1$; cold $v\!\ll\!c$ \\
\midrule
\rowcolor{white}
Axions (QCD) \\
\midrule
PQ (pseudo) Nambu–Goldstone boson; $B=0$ \\
photon coupling via suppressed $g_{a\gamma\gamma}\,a\,\mathbf E\!\cdot\!\mathbf B$ \\
Minimally coupled scalar field \\
Cold production; negligible self-interaction \\
\midrule
PBH \\
\midrule
Early-universe collapse constraint \\
Astrophysically neutral \\
Black-hole solutions of GR \\
Rare encounters \\
\bottomrule
\rowcolor{white}
\\
\\
\end{tabularx}
\end{minipage}
\begin{minipage}{0.51\textwidth}
\raggedright
\includegraphics[width=\linewidth]{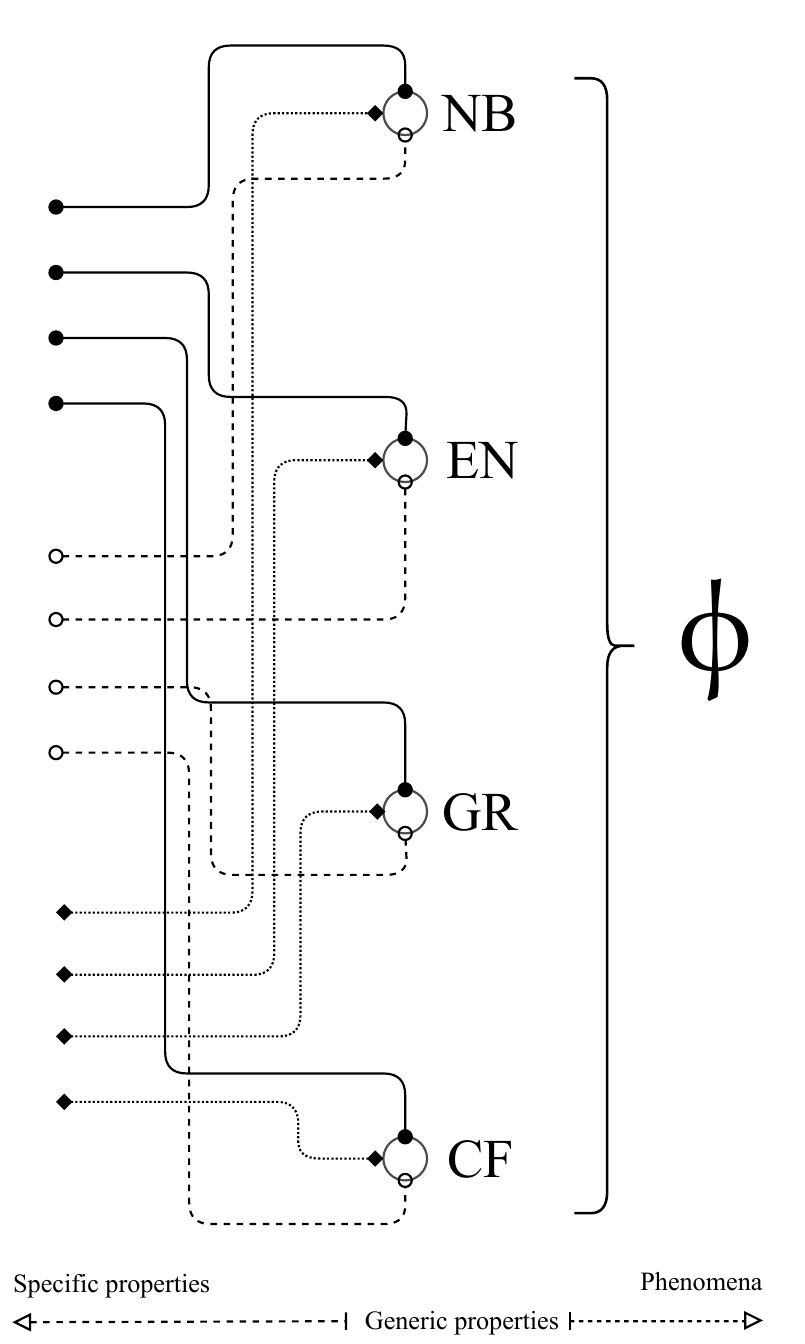}
\end{minipage}

\caption{Determinates (left) for determinables (right) which act as proxy explanations for phenomena.}
\label{fig:vdm-side-by-side}
\end{figure}

In short, dark matter realism fails here not because determinables are
false, but because we lack hyperintensional access to their realizers;
without that grain, causal--descriptive reference and natural-kind
individuation cannot be secured. Thus, while determinables guide
modeling, they bear no realist weight on a causal--descriptive semantics
absent access to their determinates; realism here must wait for
truth-makers or retreat. This embodies the conclusions of Martens
(\citeproc{ref-martens2022dark}{2022}) and Allzén
(\citeproc{ref-allzen2021scientific}{2021}): Martens
(\citeproc{ref-martens2022dark}{2022}) argues that dark matter realism
requires a descriptive semantics that is neither empirically
underdetermined nor vacuously true --- which Allzén
(\citeproc{ref-allzen2021scientific}{2021}) argues is precisely the
descriptive content that canonical forms of empirical confirmation via
detection provide. One can remain a dark-matter realist in a pragmatic
or provisional sense, but not by invoking classical scientific-realist
justification. The failure arises as a consequence of traditional
realism in combination with dark matter lacking the required empirical
flavor, not from dark matter per se. Insisting that scientific realism
at this stage vindicates dark-matter realism constitutes justificatory
overreach. For realism, empirical confirmation serves a dual purpose: it
confirms the existence of an entity and simultaneously furnishes the
semantic description of that entity with the intrinsic properties
realism requires. This exposes scientific realism --- typically applied
after the moment of confirmation --- to the epistemic reality of
twenty-first-century science. The situation is revealing: it shows that
scientific realism depends, semantically and epistemically, on a form of
empirical confirmation often unavailable in frontier physics.

\subsection{Archaic scientific realism: adapt or
perish}\label{archaic-scientific-realism-adapt-or-perish}

\noindent The issues outlined above are not unique to dark matter. They
are likely to be characteristic of scientific realism in contemporary
science, particularly in fields like cosmology, astrophysics, and
particle physics, where empirical data are often scarce or indirect ---
able to fix determinables where realism requires determinates. Modern
measurements and large-scale surveys are often summaries and integrated
effects: power spectra, lensing maps, event rates. By their nature, such
observables are compatible with many different underlying mechanisms.
Empirical confirmation of the particle-physics variety provides exactly
the semantic detail that the causal--descriptive theory of reference is
meant to capture. When ostensive definitions (``that is gold'') are
unavailable, the only substitute for sufficiently fine-grained semantics
is empirical detection:

\begin{quote}
But Locke's old problem, the \emph{qua} problem, looms. A chunk of gold,
for example, would have to be included in the reference of the term
`gold' in virtue of one underlying property and in the reference of
`metal' in virtue of quite another. Investigating the world can only
tell us what structural properties objects have. It cannot divulge which
ones are of interest in specifying the reference of particular terms. To
be sure, once we know a lot of physics and chemistry, when we can
actually specify the respects in which the internal structures of all
samples of gold must agree -- or, in other words, when we can provide an
explicit description of the internal structure -- there is no problem.
Prior to that point, what is required is a way of picking out which
structural feature is implicated in reference for a particular natural
kind term. (\citeproc{ref-stanford2000refining}{Stanford \& Kitcher,
2000, p. 109})
\end{quote}

\noindent In other words, without a principled way to pick out which
underlying property (the `qua' aspect) is doing the referential work for
`dark matter', our semantic success remains in limbo. Scientific
realism, as traditionally formulated, assumes that such reference-fixing
properties will eventually be revealed by investigation. But in cases
like dark matter, we have the phenomena and a label --- `dark matter'
--- without a determinate property to anchor that label. This reinforces
the notion that realism here must either wait for an empirical
breakthrough or revise its expectations.\footnote{Some philosophers
  equipped with both foresight and providence have laid the conceptual
  foundations to address precisely such a situation: Dawid
  (\citeproc{ref-dawid2013string}{2013}); Dawid
  (\citeproc{ref-dawid2003realism}{2003}); Dawid et al.
  (\citeproc{ref-dawid2015no}{2015}); Dawid
  (\citeproc{ref-dawid2017bayesian}{2017}); Dawid
  (\citeproc{ref-dawid2019significance}{2019}); McCoy
  (\citeproc{ref-MCCOY202115}{2021}); McCoy
  (\citeproc{ref-mccoy2019epistemic}{2019}), Vickers
  (\citeproc{ref-vickers2023future}{2023}), Wolf \& Read
  (\citeproc{ref-wolf2025navigatingpermanentunderdeterminationdark}{2025}).}
Scientific realism has stayed admirably close to the epistemic practices
of twentieth-century physics, so it is unsurprising that the scope of
its realism extends exactly to the boundary of those practices. Realism
--- born in the ``golden age'' of particle physics, where canonical
experiments reliably confirmed the entire Standard Model --- struggles
to accommodate theoretical entities like dark matter, whose confirmation
does not follow the usual experimental paradigm. Plausibly, the
dark-matter case is not an anomaly but a representative instance of the
epistemic state of contemporary physics, which operates at scales,
regimes, and physical boundaries practically (and sometimes in
principle) inaccessible to the standards of twentieth-century
physics.\footnote{The situation is succinctly summarized by theoretical
  physicist Leonard Susskind: ``The physicist's guiding star has always
  been experimental data, but in this respect things are harder than
  ever. All of us (physicists) are very aware of the fact that
  experiments designed to probe ever deeper into the structure of matter
  are becoming far bigger, more difficult, and costlier. The entire
  world's economy for one hundred years would not be nearly enough to
  build an accelerator that could penetrate to the Planck scale. Based
  on today's accelerator technology, we would need an accelerator that's
  at least the size of the entire galaxy!'' Susskind
  (\citeproc{ref-Susskind2006tcl}{2006})}

If this pattern \emph{is} endemic --- not a temporary technological
impasse but a stable epistemic and theoretical reconfiguration of
scientific inquiry --- then classical realism risks obsolescence as a
doctrine of realism for much of contemporary and future science.
According to this diagnosis, the existential threat to scientific
realism is not external, but internal. It does not come by
counterexample from a determined anti-realist, but from scientific
realism's own internal structure being unable to accommodate the
evolution of scientific theorizing and epistemic practices. Unless it
addresses these issues, scientific realism risks withering away in a
changed scientific landscape, barren of the once-abundant forms of
empirical confirmation. The choice is to either insist on classical
standards and suspend realism, or modify our conceptual framework of
scientific realism to fit twenty-first-century science. The wager is
clear: either the evidence will, at some point, resolve into the right
kind of grain, or the philosophy must.

\section{References}\label{references}

\phantomsection\label{refs}

\end{document}